\documentclass[iop]{emulateapj}

\usepackage{natbib}
\usepackage{graphicx}
\usepackage{bbm}
\usepackage{amsmath}
\usepackage{amssymb}
\usepackage{color}

\newcommand{\beq}{\begin{equation}}
\newcommand{\eeq}{\end{equation}}
\newcommand{\be}{\begin{equation}}
\newcommand{\ee}{\end{equation}}

\newcommand{\muG}{\ensuremath{\mu}\textrm{G} }

\begin{document}

\title{The Galactic Magnetic Field}
\author{Ronnie Jansson and Glennys R. Farrar}
\affil{Center for Cosmology and Particle Physics,
Department of Physics\\
New York University,
New York, NY 10003, USA}


\begin{abstract}
With this Letter, we complete our model of the Galactic magnetic field (GMF), by using the WMAP7 22 GHz total synchrotron intensity map and our earlier results to obtain a 13-parameter model of the Galactic random field, and to determine the strength of the striated random field.  In combination with our 22-parameter description of the regular GMF, we obtain a very good fit to more than forty thousand extragalactic Faraday Rotation Measures (RMs) and the WMAP7 22 GHz polarized and total intensity synchrotron emission maps.   The data calls for a striated component to the random field whose orientation is aligned with the regular field, having zero mean and rms strength $\approx$20\% larger than the regular field.    A noteworthy feature of the new model is that the regular field has a significant out-of-plane component, which had not been considered earlier.  The new GMF model gives a much better description of the totality of data than previous models in the literature.
\end{abstract}

\section{Introduction} 

The magnetic field of the Galaxy (GMF) eludes our understanding in many respects, in spite of a great deal of effort measuring and modeling its various components.  The most familiar components of the GMF are the large-scale regular fields and the small-scale random fields, reviewed in  \citet{beuermann+85} and \citet{Sun:2008}.  The random fields are due to a number of phenomena including supernovae and other outflows, possibly compounded by hydrodynamic turbulence, which are expected to result in randomly-oriented fields with a coherence length $\lambda$ of order 100 pc or less \citep{Gaensler:1995, Haverkorn:2008}.   In addition to these, there may be ``striated" random fields whose orientation is aligned over a large scale, but whose strength and sign varies on a small scale.  Such striated fields can be produced by differential rotation of a medium containing small scale random fields, the levitation of bubbles of hot plasma carrying trapped randomly oriented fields away from the disk, or both.   
Striated fields are a special case of the more generic possibility of anisotropic random fields introduced in \citet{Sokoloff:1998}, which can be considered a superposition of  multiple striated and purely random fields. 

In \citet{jf12} (JF12 below) we introduced a 22-parameter model of the large-scale, coherent or regular Galactic Magnetic Field, constrained by a simultaneous fit to the WMAP7 Galactic Polarized Synchrotron Emission (Stokes $Q$ and $U$, or, collectively, PI) maps \citep{WMAP7} and more than forty thousand extragalactic Faraday rotation measures (RMs). The model includes an out-of-plane component and takes into account the possible presence of a striated random field and/or need for rescaling the assumed density of relativistic electrons.  The new model gives a much better accounting of the observational data than achieved earlier.   Key to the feasibility of the undertaking is the fact that the different data types probe different aspects of the field and are sensitive to different sources of uncertainty.   The Faraday rotation measure is the integrated line-of-sight component of $\vec{B}$, weighted by the thermal electron density, for which we adopt the standard NE2001 \citet{Cordes:2002} model.  The polarized and total synchrotron intensity are also line-of-sight integrals, but sensitive to the transverse rather than longitudinal component of the GMF and weighted by the relativistic (cosmic ray) electron density, $n_{\rm cre}$.

We present here a 13-parameter model of the spatial variation of the strength of the Galactic random field (GRF), obtained by applying the methodology and infrastructure of JF12 to the total, unpolarized Galactic synchrotron emission from WMAP.  The new random-field model provides a good fit to the large-scale structure of the total synchrotron emission $I$ of the galaxy.  This approach determines the rms value of $B_{\rm rand}$ as a function of position within the Galaxy but is not sensitive to the ``internal structure" of the random field, i.e., maximum coherence length and power-spectrum.  Combining these results with data on the variance can be used to constrain the distribution of coherence lengths, as will be presented elsewhere.

A global parameterization of the GMF including both coherent and random fields is needed for many practical purposes.  One important application is predicting the pattern of ultra-high energy cosmic ray (UHECR) deflections, to identify sources and gain insight into composition; an early effort to characterize the Galactic random field for UHECR studies was made by \citet{Tinyakov:2005}.  Another important application is to understand the spatial variation and anisotropy of the diffusion coefficients for Galactic cosmic ray propagation, needed to obtain more realistic predictions for the diffuse gamma ray background and Galactic cosmic ray spectra and thus to have more accurate models for the backgrounds to possible Dark Matter signals.

\section{Model Field}

We allow for three distinct types of magnetic structures in our analysis: large-scale regular fields, striated fields, and small-scale random fields.  These different structures can be disentangled because they contribute in general to different observables:  the large-scale regular field can contribute to all the observables, $I$, $PI$ and RM;  the striated field can contribute to $I$ and $PI$ but not to RMs, in leading order, due to its changing sign; while the purely random field contributes only to the total synchrotron emission, $I$.  

In JF12 we considered two forms of $n_{\rm cre}$ and adopted the GALPROP model provided by A. Strong (private communication, 2009), allowing for a possible overall rescaling by a factor $\alpha$.  Increasing the scale height and rescaling the GALPROP $n_{\rm cre}$, as could arise if anisotropic diffusion due to the out-of-plane field were incorporated in GALPROP, further improves the fit;  \citet{Strong+CReSpec11} and \citet{Jaffe+MNRAS11} have also shown evidence of a need for a modification in the intersteller electron spectrum.  In future work we intend to self-consistently solve for the GMF and $n_{\rm cre}$ simultaneously, but for the present we adopt  JF12's partial-improvement of optimizing the normalization but not the shape of the GALPROP $n_{\rm cre}$.   

The analysis of JF12 describes the coherent field as a superposition of disk, toroidal halo, and ``X-field" components, with a total of 21 parameters.  Each of these field components was initially allowed to have a separate amount of striation, with $B^2_{d,h,X \, \rm stri}\beta_{d,h,X} \, B_{d,h, X \,\rm reg}^2$ and the orientation of a given component of the striated field taken to be aligned with corresponding component of the local coherent field.  This introduces 3 more parameters than used to describe the coherent field alone.  However there was not a significant improvement in the fit using the freedom of separate $\beta_{d,h,X} $, so a single $\beta$ value was used for all components in the final parameter optimization.  When there is a single $\beta$ for all components, it means that the possible striated field is aligned with the total local regular field resulting in a degeneracy between the strength of the striated component of the magnetic field and the relativistic electron density:  if we write the multiplicative factor as $\gamma=\alpha(1+\beta)$, we can interpret $\alpha$ as being a rescaling factor of the nominal relativistic electron density, with $B^2_{\rm stri}=\beta B_{\rm reg}^2$.  
The combined fit to $Q$, $U$, and RMs in JF12 established that $\gamma = 2.92 \pm 0.14$;  our present fit to $I$ allows us to break the degeneracy and determine $\alpha$ and $\beta$ separately.  A future improvement will be to allow $\alpha$ and $\beta$ to be spatially varying.

After some experimentation, to get a good fit to the observations with a minimum of model complexity, we found that a good fit to the WMAP7 $I$ map can be obtained by taking the random field to be a superposition of a disk component, with a central region and eight spiral arms having the same geometry as for the coherent field, plus an extended random halo component.  The total random field has an rms strength of $B_{\rm rand}=\sqrt{B_{\rm disk}^2+B_{\rm halo}^2}$. 

The disk component of the GRF is modeled as the product of a radial factor times a vertical profile, taken to be a Gaussian of width $z_0^{\rm disk}$.  In the inner 5 kpc, the radial factor is a constant: $b_{\rm int}$.  Starting at 5 kpc, the rms strength is different from one arm to another.  Its value at 5 kpc in the $i$th spiral arm is denoted $b_i$, and it decreases as $\sim 1/r$ at larger radii. The extended random halo field is simply  the product of an exponential in the radius and a Gaussian in the vertical direction, respectively:  $
B_{\rm halo}=B_0 \, exp[-r/r_0] \, exp[-z^2/2 z_0^2]$.

\section{Method and Data}\label{method}
Our approach to constraining the parameters of the GMF is described in detail in JF12 but we review it briefly here, since the present analysis uses the same methodology.  
The observables used in the present analysis are the average values of $I$ in 13.4 square degree HEALpix cells\citep{Gorski:2005} and we minimize $\chi_I^2=\sum_i (I_{\rm data,i}-I_{\rm model,i})^2 / \sigma^2_{I,i}$ as a function of the Galactic Random Field (GRF) parameters, taking the  JF12 large scale GMF parameters and value of $\gamma$ as given.  The conceptual foundation on which this and the JF12 project is based, is the recognition that the $\sigma_i$'s 
are themselves measurable observables.  The variances in the primary observables are not merely due to observational uncertainties, but include and are dominated by the \emph{astrophysical} variance caused by turbulent magnetic fields, inhomogeneities in the interstellar medium, and source contributions to the RMs.   We {\em measure} $\sigma_i$ for a given observable, from the variance in values of that observable in the 16 sub-pixels of the $i$th 13.4 square degree HEALPix pixel.  

\begin{figure}
\centering
\includegraphics[width=0.9\linewidth]{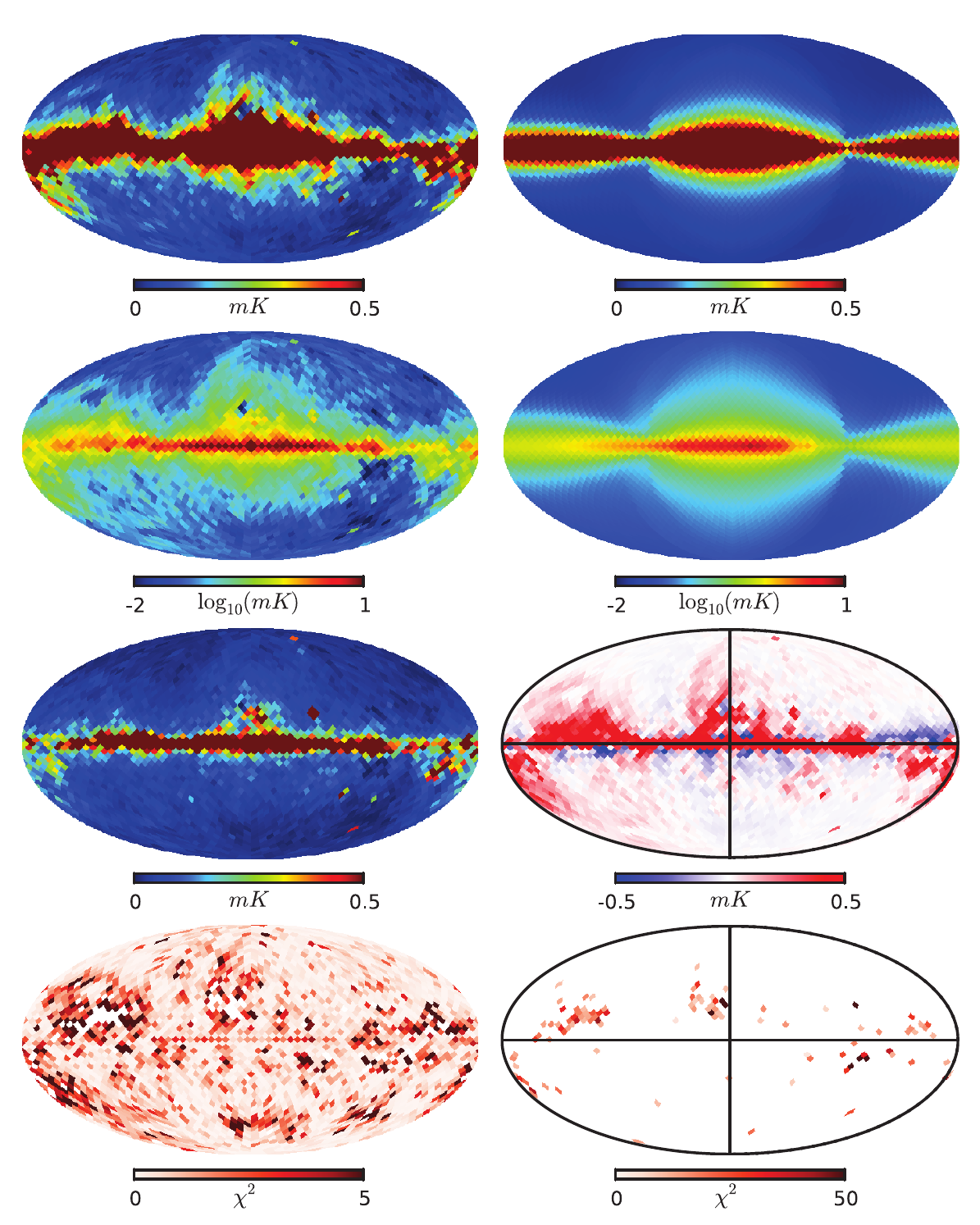}
\caption{\emph{Top left:} WMAP 22 GHz unpolarized synchrotron radiation. \emph{Top right:} Simulated 22 GHz synchrotron radiation for the optimized model. \emph{Second row:} As top row, in logarithmic scale. \emph{Third row, left panel:} The estimated $\sigma$ of the unpolarized synchrotron data. \emph{Third row, right panel:} The difference between observed and simulated data. \emph{Bottom left:} The $\chi^2$ of the fit, with 101 pixels removed (see section \ref{method}). \emph{Bottom right:} The $\chi^2$ of the masked-out pixels.  Skymaps are in the Mollweide projection with galactic longitude zero in the center, increasing to the left. }\label{randomGMFSkymaps}
\end{figure}

It is possible to estimate the unpolarized synchrotron emission without implementing a fully stochastic random field, which is computationally expensive, by generalizing a trick introduced in \cite{jf12} for a striated random field.  Assuming a spectral index of 3 for the relativistic electrons, the synchrotron emissivity of a volume element in a region with local field $\vec{B}$  is  $ \propto n_{\rm cre} \, B_\perp^2 = n_{\rm cre} \, B^2 {\rm sin}^2\theta$ where $\theta$ is the angle between the line of sight and the direction of $\vec{B}$.  If we add a random field of rms strength $B_{\rm rand}$ to $\vec{B}_{\rm reg}$ and average over the direction the random field may take, represented by $<É>$, then 
$< B_\perp^2 > = B_{\rm reg}^2 \, {\rm sin}^2 \theta + \frac{2}{3} \, B_{\rm rand}^2  $.
To simulate the total synchrotron emission from a superposition of random, striated and regular fields, without ever creating realizations of the random field, we use Hammurabi \citep{Waelkens:2009} with the model $n_{\rm cre}$, but replacing
\begin{equation}
B_{\rm reg}^2 \rightarrow \alpha \, (1+\beta) \,B_{\rm reg, model}^2 \, \left(1+ \frac{2}{3}\frac{B_{\rm rand}^2}{ (1+\beta) B_{\rm reg,model}^2 \sin^2\theta}\right), 
\end{equation} 
with $\alpha \, (1 + \beta) = \gamma $ fixed from JF12.  Since $\alpha$ and $1 + \beta$ enter the expression for the total intensity differently, fitting the total intensity data allows them to be separately determined.   As in  JF12, we use a Metropolis Markov Chain Monte Carlo (MCMC) algorithm to find best-fit parameters and confidence levels.

\begin{figure}
\centering
\includegraphics[width=1\linewidth]{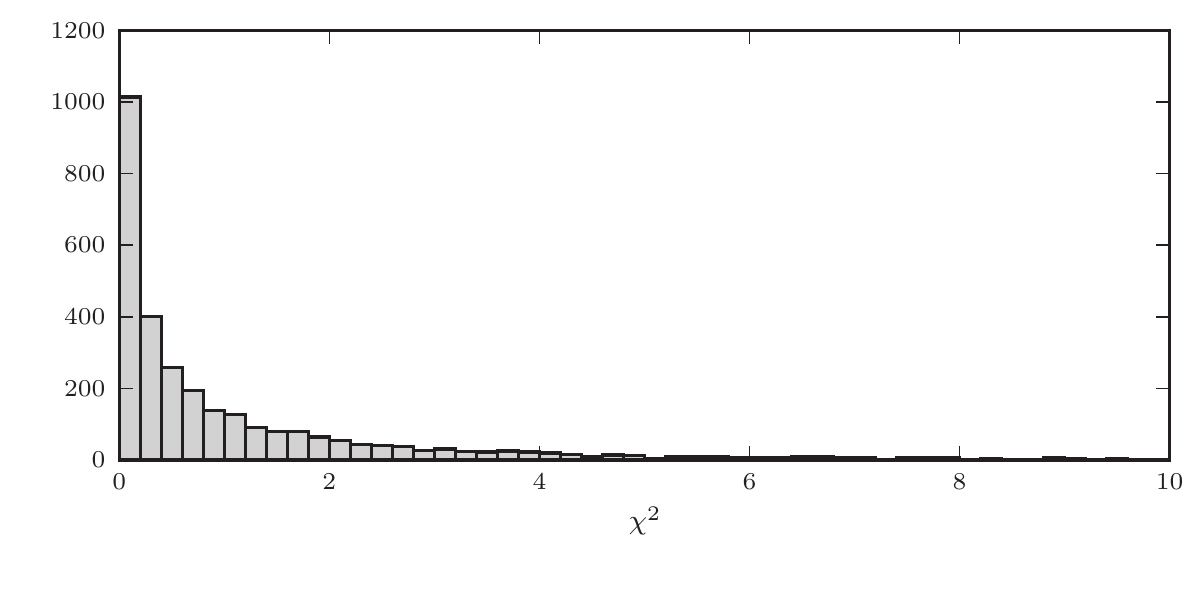}
\caption{A histogram of $\chi^2$ for the 3072 pixels covering the sky. To reduce the impact of mainly nearby structures the 101 pixels with $\chi^2$ greater than 10 are removed from the data, and a second optimization is performed to obtain the best-fit parameters presented in Table \ref{tab:para}.  }
\label{hist}
\end{figure}

The WMAP total Galactic synchrotron intensity map used for the present analysis is shown in linear and logarithmic scales in the top two rows of the left-hand column in Fig. \ref{randomGMFSkymaps}, and the map of measured $\sigma$ values for each 13.4 square degree HEALPix pixel is shown in the left panel of the third row.  
In order that fluctuations due to nearby strong individual contributions do not distort the determination of the global properties of the Galactic random field, after performing an initial fit for the model parameters we mask-out the pixels with $\chi_i^2 > 10$, and then perform a second (and final) parameter optimization.  The distribution of $\chi_i^2$ values from the initial model fit is shown in Fig. \ref{hist}.  The lower panels of Fig. \ref{randomGMFSkymaps} show the $\chi_i^2$ per pixel maps (left) with the masked pixels removed and (right) showing the 101 masked-out pixels;  the $\chi_i^2$'s shown in these maps are from the second model optimization.  

\begin{table}
\caption{Best-fit parameters of the random field, with $1-\sigma$ intervals.}
\begin{tabular}{lll}
\tableline
Field &  Best fit Parameters &  Description  \\
\tableline
Disk            & $b_1= 10.81\pm 2.33 \,\muG$                 &   field strengths at $r=5$ kpc \\
component       & $b_2= 6.96 \pm 1.58 \,\muG$                 &                                \\
                & $b_3= 9.59 \pm1.10 \,\muG$                 &                                \\
                & $b_4= 6.96\pm0.87 \,\muG$                 &                                \\
                & $b_5= 1.96\pm1.32 \,\muG$                 &                                \\
                & $b_6=16.34\pm2.53 \,\muG$                 &                                \\
                & $b_7=37.29\pm2.39 \,\muG$                 &                                \\
                & $b_8= 10.35\pm4.43 \,\muG$                 &                                \\
                & $b_{\rm int}=7.63\pm 1.39\,\muG$          &   field strength at $r<$ 5 kpc \\
                & $z^{\rm disk}_0=0.61\pm0.04$ kpc        &   Gaussian scale height of disk\\
\tableline  
Halo          & $B_0 = 4.68\pm1.39 \,\muG$              &   field strength         \\
component       & $r_0 =10.97\pm3.80 $ kpc             &   exponential scale length   \\
                & $z_{\rm 0} = 2.84\pm1.30$ kpc           &   Gaussian scale height     \\
\tableline
Striation       & $\beta = 1.36 \pm 0.36$                 &  striated field $B^2_{\rm stri} \equiv \beta  B^2_{\rm reg}$     \\
\tableline

\end{tabular}\label{tab:para}
\end{table}

\begin{figure}
\centering
\includegraphics[width=1\linewidth]{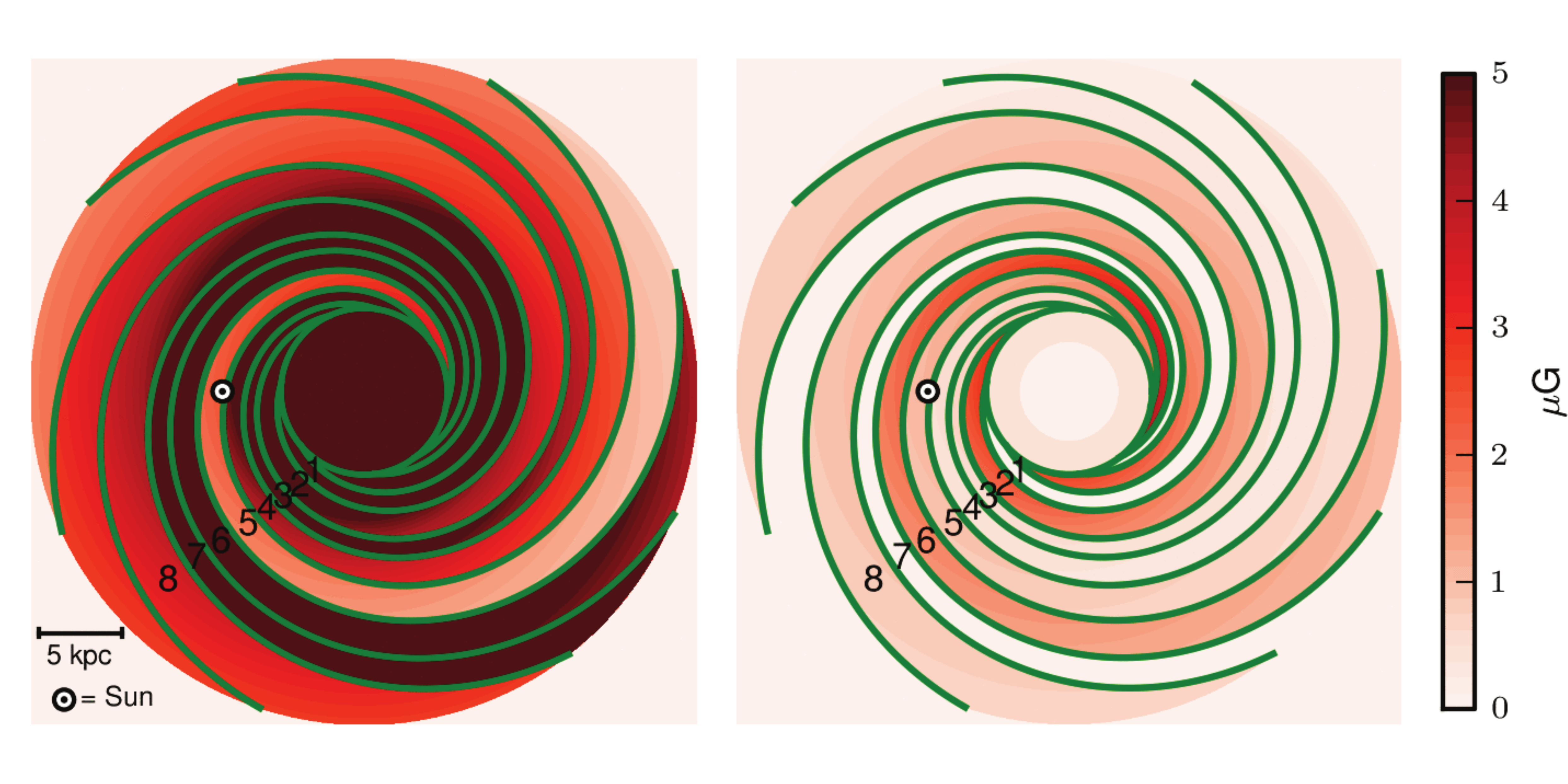}
\caption{\emph{Left panel:} The random field in the disk.   \emph{Right panel:} The disk component of the  JF12 coherent field model for comparison;  it is clockwise in rings 3-6 and counterclockwise in 1,2,7, and 8. }\label{B_disk_random}
\end{figure} 


\section{Results of the Fit}

The best-fit parameters for this model are given in Table \ref{tab:para}.  After finding the minimum chi-squared, an additional 100k MCMC iterations were performed to obtain a probability distribution in parameter space, reflecting the extent to which the observations do a better job constraining some parameters than others;  the errors given in Table \ref{tab:para} are obtained from that distribution, by marginalizing over the other parameters.  

The top two rows of Fig. \ref{randomGMFSkymaps} show the comparison between data and predicted synchrotron intensity for this best-fit random field.  The right-hand panel in the third row shows the residual between the WMAP 22-GHz total synchrotron emission map and that produced by the random field model.  It is concentrated near the Galactic plane and in isolated other regions, and is compatible with being due to individual sources such as Cen A, the North Polar Spur and other nearby structures not modeled here.  The magnitude of the residual between model and data is small compared to the total signal, as most easily seen from the left panel of the second row, where the total intensity is shown in a logarithmic scale.  

The 13-parameter form of the Galactic random field adopted here proves to be sufficiently general to give a very good accounting of the data.  In combination with the  JF12 coherent field and the striated random field, it provides an excellent fit to the total intensity, $I$, with a reduced $\chi^2$ ($\chi^2$ per degree of freedom, $\chi_{\rm dof}^2$) after the second optimization of 1.064 with 2957 degrees of freedom.  (One should not attach too much significance to the exact value of $\chi_{\rm dof}^2$ for the GRF model, since it depends on the arbitrary cut used to remove pixels with large ``individualistic" contributions;  as evident from the map of residuals in Fig. \ref{randomGMFSkymaps}, additional pixels could be placed in this category which would decrease $\chi_{\rm dof}^2$.)  As noted, the fit to $I$ breaks the degeneracy between rescaling $n_{\rm cre}$ and the presence of a  striated random field.

The left panel of Fig. \ref{B_disk_random} shows the disk component of the random field, with the magnitude of the coherent disk field from JF12 in the right panel for comparison.   The average rms strength of the disk component of the random field at the solar circle is 6.6 $\mu$G, but it varies strongly from arm to arm.  The spiral-arm model itself should not be taken too literally -- it is presumably no more than a simplified encoding of some important features of the structure.   

Due to the large value of the random and striated fields compared to the coherent field, we cannot predict the value of the field at a particular position since the fluctuations dominate the mean value.  Nonetheless, we should check whether the predicted range of values for the total field in the solar neighborhood is consistent with observations -- keeping in mind that the rms component has O(1) variance locally, so the actual local field at any particular position can be expected to differ significantly from the result of combining the local coherent, striated and random components in quadrature.  The estimate is additionally uncertain due to our position near the boundary between the 4th and 5th arms, since the simple arm geometry assumed in the present models is only expected to be valid in some average sense and may be locally modified, and the parameters specifying the geometry of the arms in  JF12 were taken from the NE2001 model of $n_e$ rather than being free parameters of the GMF model.   With those caveats, the vertical component of the local coherent field is 0.2 $\mu$G and the horizontal component is 0.5-1.2 $\mu$G in the 4th and 5th arms.  Combining the coherent, striated and random components in quadrature gives an estimate of the magnitude of the local field of 3-5 $\mu$G, with the range reflecting the values obtained for the 5th and 4th arms.  Given the O(1) variance from the random field, this estimated local GMF value is consistent with the 6 $\mu$G value commonly cited \citep{beckG&EGMF08}.  The more recent studies of \citet{Taylor:2009} and \citet{Mao:2010} are complementary, with the former having larger sky coverage and the latter a higher density of well-measured extragalactic sources but restricted to the polar caps.  Our results are consistent with both, within the observational uncertainties and predicted fluctuations.  For instance, \citet{Mao:2010} finds that the local random field is larger than the local coherent field, as we do, and estimates the halo random field above the solar system based on the variance in RMs in the polar caps to be $\approx 1\, \mu$G in some average sense, consistent with our fit which decreases slowly from $\approx 2\, \mu$G in the Galactic plane, and is $1 \, \mu$G  about 3 kpc above the plane.

\section{Summary}

We have presented here a 13-parameter model of the random component of the Galactic magnetic field, to complete our characterization of the Galactic magnetic field. Taken together, our comprehensive GMF model fit utilizes 36 parameters, including the 21-parameter JF12 model for the coherent field, the strength of a striated random field proportional to and aligned with the local coherent field, and an overall rescaling of the GALPROP cosmic ray electron density.  This model provides an excellent fit to more than 40,000 extragalactic RMs and the WMAP polarized and unpolarized Galactic synchrotron emission maps -- a total of nearly 10,000 independent observables, each with a measured variance.  The striated random field has vanishing mean and is locally aligned with the regular field; its energy density is found to be $\approx 35$\% larger than that of the regular field.  The fit indicates that the GALPROP cosmic ray electron density provided by A. Strong in 2009 is $\approx 20$\% too low in the relevant energy regime, and/or its scale height may need to be revised; indeed, direct comparison to the current observations shows that including the correction factor inferred by our method gives a better fit in the relevant energy region.   

The GMF model provided here and in JF12 is a substantial improvement over the previous state-of-the-art GMF models in several ways.  First, the coherent component gives a much better fit to the RM, $Q$ and $U$ data than any other model.  The JF12 coherent field model and $\gamma$ rescaling has a $\chi^2$ per degree of freedom of 1.096 for the 6605 observables (pixels of RM, $Q$ and $U$) while the fit is much worse for other recent models in the literature, with $\chi_{\rm dof}^2$ of  2.663 \& 4.971 for \citet{pshirkov+11} BSS \& ASS respectively and 1.672 for \citet{Sun:2008}, with their 2010 parameter update.  Second, the functional form used in our model is better and more general:  we enforce that the field is divergenceless and include out-of-plane and striated components.  The improvement due to our functional form is evident by re-optimizing the parameters of the other models with our MCMC procedure and Hammurabi; then their $\chi_{\rm dof}^2$ values become 1.452, 1.591 and 1.325 respectively -- still vastly worse than the 1.096 of JF12, considering there are nearly 6600 degrees of freedom.   Finally, we provide for the first time a detailed model of the random field strength.  The final fit to 2971 pixels of $I$ including the contributions of all three components of the GMF (coherent, striated and random) has $\chi_{\rm dof}^2 = 1.064$.  We emphasize that there are two additional sources of uncertainty which we cannot quantify:  the uncertainty in the relativistic (and thermal) electron distributions, needed to predict the RM and synchrotron emission observables for a given GMF model, and a possible inadequacy in the field model which could arise due to not having chosen a sufficiently general functional form.  Future work will aim to address both of these issues, through self-consistent determination of the electron thermal and relativistic electron distributions, and exploration of additional field models. 

The total energy in the random and striated fields in this model is roughly six times that of the coherent field.  Taking all three components together -- random, striated and regular -- we find that the magnetic energy of the Galaxy is $\approx 5 \times 10^{55}$ erg.  The local magnetic field strength is predicted to be $3-5 \mu$G, with an uncertainty of order of the rms local random field, $3 \mu$G; this is consistent with the 6 $\mu$G value of \citep{beckG&EGMF08}.  The average rms strength of the disk component of the random field at the solar circle is 6.6 $\mu$G, but has significant arm-to-arm variations.  The vertical component of the JF12 coherent local field is 0.2 $\mu$G, which is compatible with observational estimates \citep{HanQiao94,Taylor:2009,Mao:2010}.  

The improved determination of the Galactic random and coherent fields provided by this work represents a significant advance in our understanding of the Galaxy.  Explaining the Galactic magnetic field will be an important challenge to theory, and using this more exact model will allow greater fidelity in making prediction for a host of important applications, from the deflections of ultra-high energy cosmic rays to the spectra and distribution of Galactic cosmic rays.  

This research has been supported in part by NSF-PHY-0701451 and NASA grant NNX10AC96G.


\end{document}